\documentclass[twocolumn,pra,superscriptaddress,longbibliography]{revtex4-2}
\usepackage{graphicx,amsmath,amssymb}
\usepackage[colorlinks=true, citecolor=blue, urlcolor=blue, linkcolor=blue]{hyperref}

\begin{document}

	\title{ Limitations of strong coupling in non-Markovian quantum thermometry}
	
	\author{Qing-Shou Tan}
	\email{qstan@hnist.edu.cn}
	\affiliation{Key Laboratory of Hunan Province on Information Photonics and Freespace Optical Communication, College of Physics and Electronics,
		Hunan Institute of Science and Technology, Yueyang 414000, China}
	\author{Yang Liu}
	\affiliation{Key Laboratory of Hunan Province on Information Photonics and Freespace Optical Communication, College of Physics and Electronics,
		Hunan Institute of Science and Technology, Yueyang 414000, China}
	\author{Xulin Liu}
	\affiliation{Key Laboratory of Hunan Province on Information Photonics and Freespace Optical Communication, College of Physics and Electronics,
		Hunan Institute of Science and Technology, Yueyang 414000, China}
	\author{Hao Chen}
	\affiliation{Key Laboratory of Hunan Province on Information Photonics and Freespace Optical Communication, College of Physics and Electronics,
		Hunan Institute of Science and Technology, Yueyang 414000, China}
	\author{Xing Xiao}		
	\affiliation{	School of Physics and Electronic Information, Gannan Normal University, Ganzhou, Jiangxi, 341000, China
	}
	\author{Wei Wu}
	\affiliation{Key Laboratory of Theoretical Physics of Gansu Province, and Lanzhou Center for Theoretical Physics, Lanzhou University, Lanzhou, 730000, China}
	
	\date{\today }
	
	\begin{abstract}

    We investigate quantum thermometry using a single-qubit probe embedded in a non-Markovian environment, employing the numerically exact hierarchical equations of motion (HEOM) to overcome the limitations of  Born–Markov approximations. Through a systematic analysis of the dynamical and steady-state behavior of the quantum signal-to-noise ratio (QSNR) for temperature estimation, we identify several key findings that challenge the conventional expectation that strong coupling necessarily enhances thermometric performance. In non-equilibrium dynamical thermometry, weak system-environment coupling generally yields the optimal QSNR, whereas in the steady-state regime, strong coupling enhances sensitivity only in the ultra-low-temperature limit, while weak coupling significantly improves precision at moderately low temperatures. To optimize performance across coupling regimes, we develop a hybrid computational framework that integrates HEOM with quantum-enhanced particle swarm optimization, enabling precise quantum dynamical control under varying coupling strengths. Our results reveal fundamental constraints and opportunities in quantum thermometry, offering practical strategies for the design of high-performance quantum thermometers operating in realistic open quantum systems.

	\end{abstract}
	
	\maketitle
	
	\section{introduction}
	
	Quantum thermometry, which leverages quantum systems to estimate temperature with precision surpassing classical limits~\cite{PhysRevLett.114.220405,Paris_2016,Mehboudi_2019,PhysRevLett.127.190402,PRXQuantum.2.020322,PhysRevLett.119.090603}, has emerged as a cornerstone of quantum metrology. Its applications span a broad spectrum—from quantum computing and condensed matter physics to biological diagnostics and nonequilibrium thermodynamics~\cite{RevModPhys.78.217}. By harnessing uniquely quantum resources such as coherence~\cite{PhysRevResearch.5.043184,Smirne_2019}, criticality~\cite{PhysRevResearch.6.043094}, and entanglement~\cite{PhysRevA.110.032605,PhysRevA.101.032112,PhysRevA.109.042417}, quantum thermometry enables unprecedented sensitivity, particularly at the nanoscale, where classical strategies often fail~\cite{PhysRevLett.122.030403,PhysRevA.107.063317,PhysRevA.110.032611,PhysRevA.108.022608,PhysRevResearch.7.023100}.
	
	Contemporary quantum thermometry protocols generally fall into two paradigms~\cite{Mehboudi_2019}: steady-state and dynamical approaches. Steady-state thermometry relies on the thermal equilibrium between a quantum probe and its environment~\cite{DePasquale2018,Sztrikacs2024,Ullah2025}, with precision primarily dictated by the probe’s heat capacity.
    While experimentally robust and conceptually straightforward, this method faces intrinsic limitations at low temperatures, where the vanishing heat capacity imposes classical thermodynamic bounds and poses challenges for quantum thermometry in the ultracold regime. Interestingly, recent studies have shown that strong probe–environment coupling can mitigate the divergence of estimation error at low temperatures and significantly enhance precision even in the ultra-low-temperature limit~\cite{PhysRevA.96.062103,PhysRevA.108.032220,PhysRevA.111.052623,PhysRevResearch.7.023235}, thereby challenging conventional assumptions derived from weak-coupling analyses.
	
	In contrast, dynamical thermometry operates in the prethermal regime, exploiting transient quantum effects before the system equilibrates. These protocols benefit from short-time phenomena such as quantum coherence, non-Markovian memory, and ephemeral entanglement, which can push sensitivity beyond equilibrium-based limits. Among them, non-Markovianity—manifested as environment-induced memory effects and information backflow—has proven particularly promising. By sustaining quantum coherence and correlations, non-Markovian dynamics can provide significant metrological advantages~\cite{PhysRevA.103.L010601, PhysRevApplied.17.034073, PhysRevResearch.3.043039,PhysRevA.108.022608,PhysRevA.110.032611,PhysRevE.110.024132}.
    However, whether strong coupling can effectively enhance quantum thermometric precision in dynamical scenarios remains elusive. Although strong coupling is often associated with enhanced non-Markovian effects, standard theoretical tools—particularly perturbative approaches such as the Born–Markov master equation~\cite{breuer2002theory}—frequently fail to capture the full dynamical complexity of these systems, leading to conflicting predictions and obscuring the true thermometric potential of strongly coupled quantum probes. To fully uncover and harness these effects in realistic open-system settings, non-perturbative approaches are essential.
	
	To address these challenges, we employ the hierarchical equations of motion (HEOM) formalism, a numerically exact, non-perturbative method that rigorously captures open quantum dynamics across all coupling strengths and timescales~\cite{doi:10.1143/JPSJ.75.082001, PhysRevLett.104.250401,PhysRevResearch.5.013181,PhysRevA.85.062323,PhysRevA.86.012308,Lin2025}. Using HEOM, we investigate a spin probe coupled to a bosonic thermal bath, systematically exploring a broad parameter space encompassing coupling strength, bath cutoff frequency, and temperature. This unified approach enables a direct comparison of steady-state and dynamical thermometric performance across weak, intermediate, and strong coupling regimes.

    Our results reveal fundamental limitations of strong coupling in quantum thermometry, demonstrating that weak coupling generally enhances precision in dynamical temperature sensing. Notably, we observe that significant non-Markovian amplification of the quantum signal-to-noise ratio (QSNR) can emerge even under weak system-bath interactions—particularly for moderate reservoir correlation times in the proposed model. However, this enhancement is strictly bounded—in the strong-coupling regime, although significant non-Markovianity generates transient QSNR oscillations, it remains insufficient for reaching optimal sensitivity.
    In the steady-state regime, we identify a thermalization threshold beyond which increasing the coupling strength destabilizes equilibrium, driving the system into a non-Gibbsian state sustained by persistent system–bath correlations. While strong coupling exhibits limited advantages only in the ultra-low-temperature regime, weak coupling consistently outperforms it across moderate temperature ranges.
	
    To optimize thermometric precision in the relatively strong-coupling regime, we propose a hybrid quantum–classical control framework that integrates HEOM dynamics with quantum particle swarm optimization (QPSO)~\cite{1330875}. This approach iteratively tunes system parameters to maximize QSNR during the probe’s dynamical evolution, guided by HEOM-computed responses. Flexible, scalable, and robust under realistic conditions, this framework provides a powerful toolkit for precision thermometry in complex, non-Markovian environments.

    The structure of the paper is as follows. In Sec.~\ref{sec2}, we introduce the QSNR metric and establish the theoretical foundation based on the HEOM formalism. Section~\ref{sec3} presents a detailed analysis of quantum thermometry across different coupling regimes, encompassing both dynamical and steady-state protocols. This analysis highlights the potential for enhanced thermometric precision in the weak-coupling regime. In Sec.~\ref{sec4}, we propose an optimized hybrid framework that integrates HEOM dynamics with QPSO to efficiently maximize QSNR in moderately strong-coupling scenarios. Finally, we conclude in Sec.~\ref{sec5}.   In the three appendices, we provide supplementary materials on the HEOM method, the Bloch–Redfield master equation approach, and the quantification of non-Markovianity. Throughout the paper, we set $\hbar=k_B=1$ for the sake of simplicity.

	\section{Formulation}\label{sec2}

     \subsection{Quantum signal-to-noise ratio}
	
	In quantum temperature sensing protocols, information about the reservoir's temperature is imprinted onto the quantum state of a sensor through their mutual interaction. By performing suitable measurements on the sensor, one can infer the temperature of the surrounding environment.
	To quantify the ultimate precision limits of such temperature estimation, we invoke the framework of quantum parameter estimation theory. According to this formalism, the minimum achievable uncertainty in estimating the temperature is bounded by the quantum Cram\'er–Rao inequality~\cite{Rao1992}:
    $\delta T_{\rm min} = {1}/{\sqrt{n \mathcal{F}_T}},$
	where \( \delta T_{\rm min} \) denotes the root mean square error, \( n \) is the number of independent measurements, and \( \mathcal{F}_T \) is the quantum Fisher information (QFI) with respect to the temperature \( T \). The QFI characterizes how sensitively the quantum state responds to infinitesimal variations in \( T \), and thus sets the fundamental limit on estimation precision.
	
	To evaluate the temperature sensitivity of a general two-level quantum sensor, we express its state in the Bloch form:
    $ \hat{\rho} =  ( \mathbf{I} + \hat{\mathbf{S}} \cdot \hat{\boldsymbol{\sigma}} )/2,$
	where \( \hat{\mathbf{S}} \) is the Bloch vector. The QFI for temperature can then be calculated using the closed-form expression~\cite{PhysRevA.87.022337, Liu_2020}:
	\begin{equation} \label{qfi}
		\mathcal{F}_T = |\partial_T \hat{\mathbf{S}}|^2 + \frac{(\hat{\mathbf{S}} \cdot \partial_T \hat{\mathbf{S}})^2}{1 - |\hat{\mathbf{S}}|^2},
	\end{equation}
	where \( \partial_T \hat{\mathbf{S}} = \mathrm{Tr} \left( \partial_T \hat{\rho} \, \hat{\boldsymbol{\sigma}} \right) \) is the temperature derivative of the Bloch vector. In the special case of pure states, where \( |\hat{\mathbf{S}}| = 1 \), the QFI simplifies to \( \mathcal{F}_T = |\partial_T \hat{\mathbf{S}}|^2 \). This formulation offers an experimentally measurable approach to QFI through Bloch vector dynamics.
	
	To further characterize the sensing performance, we introduce the QSNR, defined as:
	\begin{equation} \label{qsnr}
		\mathcal{Q}_T = T^2 \mathcal{F}_T,
	\end{equation}
	which yields the relative temperature uncertainty as:
    $\delta T_{\rm min}/{T} = {1}/{\sqrt{n \mathcal{Q}_T}}$.
	A higher QSNR implies a lower relative error, making \( \mathcal{Q}_T \) a crucial figure of merit for quantum thermometry.
     Importantly, $\mathcal{Q}_T$ represents the fundamental limit of minimum relative error achievable by an optimal quantum thermometer.
	In practice, this requires measuring the expectation values of the three Pauli operators $\hat{\mathbf{S}}$, from which temperature information can be extracted through state tomography. However, due to the non-commutativity of the Pauli operators, any projective measurement would disturb the quantum state, necessitating reinitialization and repetition of the dynamical process, which makes it particularly challenging to identify the optimal measurement scheme in the presence of complex noise environments.
	Therefore, to assess the impact of the system–environment coupling strength on the thermometer's performance, we primarily aim to theoretically   explore the values of $\mathcal{Q}_T$.

    Notably, the precise evaluation of the QSNR in Eq.~(\ref{qsnr}) requires computation of both the Bloch vector $\hat{\mathbf{S}}$ and its parametric gradient $\partial_T\hat{\mathbf{S}}$, which in turn demands an exact solution for the system's time-dependent density operator $\hat{\rho}(t)$.
    Critically, $\hat{\rho}(t)$ depends on the system-bath coupling strength, its functional form, and the environmental noise spectrum. Even for the extensively studied spin-boson model, current approaches for determining $\hat{\rho}(t)$ generally require the assumption of weak system-bath coupling, employing approximations like the Born-Markov approximation, secular approximation, and polaron transformation to obtain solvable master equations~\cite{RevModPhys.89.015001}. While these approaches offer distinct advantages, all suffer from inherent limitations, failing to fully capture the model's rich dynamical behavior. 
		
    Herein, we leverage the HEOM formalism to systematically explore the impact  of  coupling strength on the QSNR and quantum thermometry performance under both transient and steady-state conditions.

   \subsection{Exact quantum dynamics of qubits via the HEOM approach}
	
	We consider a two-level quantum system (probe) coupled to a bosonic thermal reservoir at temperature \( T \). The total Hamiltonian reads:
	\begin{equation}\label{ham0}
		\hat{H} = \hat{H}_S + \hat{S} \sum_k \left(g_k \hat{a}_k^\dagger + g_k^* \hat{a}_k\right) + \sum_k \omega_k \hat{a}_k^\dagger \hat{a}_k,
	\end{equation}
	where the system Hamiltonian is given by \( \hat{H}_S = \frac{1}{2}\omega_0 \hat{\sigma}_z \), with \( \omega_0 \) the qubit transition frequency. The system–bath interaction is mediated by \( \hat{S} = \hat{\sigma}_x \), enabling energy exchange between the qubit and its environment. The bosonic reservoir is modeled by harmonic oscillators with mode frequencies \( \omega_k \), annihilation and creation operators \( \hat{a}_k \), \( \hat{a}_k^\dagger \), and coupling strengths \( g_k \).

	To characterize the reservoir properties, we employ the Drude-Lorentz spectral density, a widely adopted model in open quantum systems~\cite{breuer2002theory,PhysRevLett.104.250401,PhysRevResearch.5.013181}:
	\begin{equation}
		J(\omega) = \frac{2 \lambda \omega_c \omega}{\omega^2 + \omega_c^2},
	\end{equation}
	where \( \lambda \)  is the  dissipation strength (quantifying the system-bath coupling strength), and \(\omega_c\) denotes the cutoff frequency that inversely determines the reservoir correlation time \(\tau_c \sim \omega_c^{-1}\). 

	 Correspondingly, the environmental noise correlation function, defined via \( \hat{H}_B = \sum_k \omega_k \hat{a}_k^\dagger \hat{a}_k \) and \( \hat{B} = \sum_k \left(g_k \hat{a}_k^\dagger + g_k^* \hat{a}_k\right) \), is expressed as~\cite{doi:10.1143/JPSJ.75.082001, PhysRevLett.104.250401,PhysRevResearch.5.013181}
	\begin{eqnarray}
		C(t) &=& \langle e^{i t \hat{H}_B} \hat{B} e^{-i t \hat{H}_B} \hat{B} \rangle_B \nonumber\\ &\approx& \sum_{k=0}^{N_k} c_k e^{-\nu_k t} + \sum_{k = N_k + 1}^{\infty} \frac{c_k}{\nu_k} \delta(t),
	\end{eqnarray}
	where \( \nu_k \) are the Matsubara frequencies, and \( c_k \) the corresponding coefficients. The sum over \( k \) reflects the decomposition of the correlation function into exponentially decaying terms and a delta-function remainder, enabling an efficient hierarchical treatment. 

	To simulate the reduced dynamics of the qubit beyond the Born–Markov approximation, we employ the  HEOM approach. This non-perturbative and fully quantum framework introduces a set of auxiliary density operators (ADOs), \(\hat{ \rho}_{\vec{n}} \), whose evolution obeys~\cite{doi:10.1143/JPSJ.75.082001, PhysRevLett.104.250401,PhysRevResearch.5.013181,PhysRevA.85.062323,PhysRevA.86.012308}:
	\begin{eqnarray}
		\dot{\hat{\rho}}_{\vec{n}}(t) &=&
		\left(\hat{ \mathcal{L}}_s - \vec{n} \cdot \vec{\nu} \right) \hat{\rho}_{\vec{n}}(t)  \nonumber \\
		&&+ \hat{\Phi} \sum_{k=0}^{N_k} \hat{\rho}_{\vec{n} + \vec{e}_k}(t)
		+ \sum_{k=0}^{N_k} \nu_k \hat{\Theta}_k \hat{\rho}_{\vec{n} - \vec{e}_k}(t),
	\end{eqnarray}
	where \( \vec{n} = (n_0, n_1, \ldots, n_{N_k}) \) is a multi-index labeling the hierarchy level, \( \vec{e}_k \) denotes the unit vector along the \( k \)-th direction, and \( \vec{\nu} = (\nu_0, \nu_1, \ldots, \nu_{N_k}) \) collects the Matsubara frequencies. The involved superoperators are defined as
   $\hat{\mathcal{L}}_s = -i \hat{H}_S^{\times}, 
	\hat{\Phi} = -i \hat{S}^{\times}$, and $
	\hat{\Theta}_k = -i[ \mathrm{Re}(c_k)\, \hat{S}^{\times} + i\, \mathrm{Im}(c_k)\, \hat{S}^{\circ} ],$
	with \( \hat{o}_1^{\times} \hat{o}_2 = \hat{o}_1 \hat{o}_2 - \hat{o}_2 \hat{o}_1 \) and \( \hat{o}_1^{\circ} \hat{o}_2 = \hat{o}_1 \hat{o}_2 + \hat{o}_2 \hat{o}_1 \).
	The hierarchy is initialized with $	\hat{\rho}_{\vec{n} = \vec{0}}(0) = \hat{\rho}_S(0), \quad \hat{\rho}_{\vec{n} \neq \vec{0}}(0) = 0,$
	where \( \vec{0} = (0, 0, \ldots, 0) \) is the zero vector.
	
    In numerical implementations of the HEOM method, the number of auxiliary operators and the Matsubara frequency cutoff  $N_k$
    are systematically increased to ensure convergence. Due to its non-perturbative nature and ability to account for memory effects, HEOM offers a precise and versatile framework for studying open quantum systems. It is particularly effective for investigating non-Markovian thermometry and quantum metrology protocols in regimes characterized by strong system-environment coupling and complex structured environments (see Appendix \ref{appa} for more details).
	

    \begin{figure*}[tbp]
	\centering
	\includegraphics[width=18.1cm]{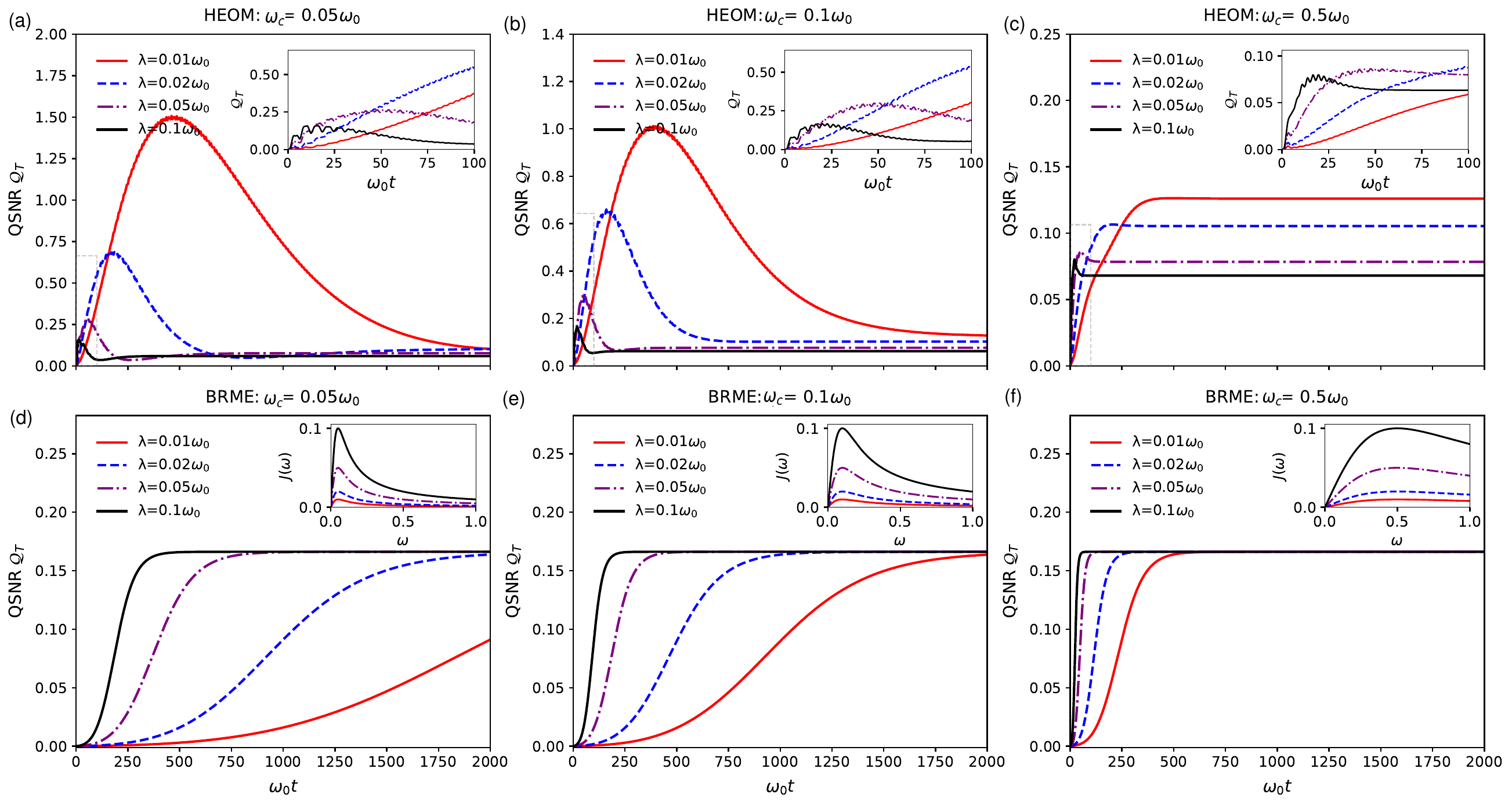}
    \caption{(a)–(c) QSNR \(\mathcal{Q}_T\) dynamics for varying coupling strengths (quantified by the dimensionless parameter \(\lambda/\omega_0\)) under different environmental parameters, computed using the HEOM method. Results correspond to \(T/\omega_0 = 0.2\) with \(\omega_c/\omega_0 = 0.05\), \(0.1\), and \(0.5\), respectively. Insets show the short-timescale \(\mathcal{Q}_T\) dynamics. (d)–(f) QSNR \(\mathcal{Q}_T\) dynamics for varying \(\lambda\) under identical environmental parameters, calculated via the Bloch-Redfield master equation (BRME). The insets depict how increasing the cutoff frequency $\omega_c$ induces spectral smoothing in $J(\omega)$, thereby suppressing environmental memory effects. Here, the system is initialized in the equal superposition state $|\psi(0)\rangle = (|0\rangle + |1\rangle)/\sqrt{2}$ of the two eigenstates of $\hat{H}_S$.
    }\label{fig1}
    \end{figure*}

\begin{figure}[bh]
	\centering\includegraphics[width=8.5cm]{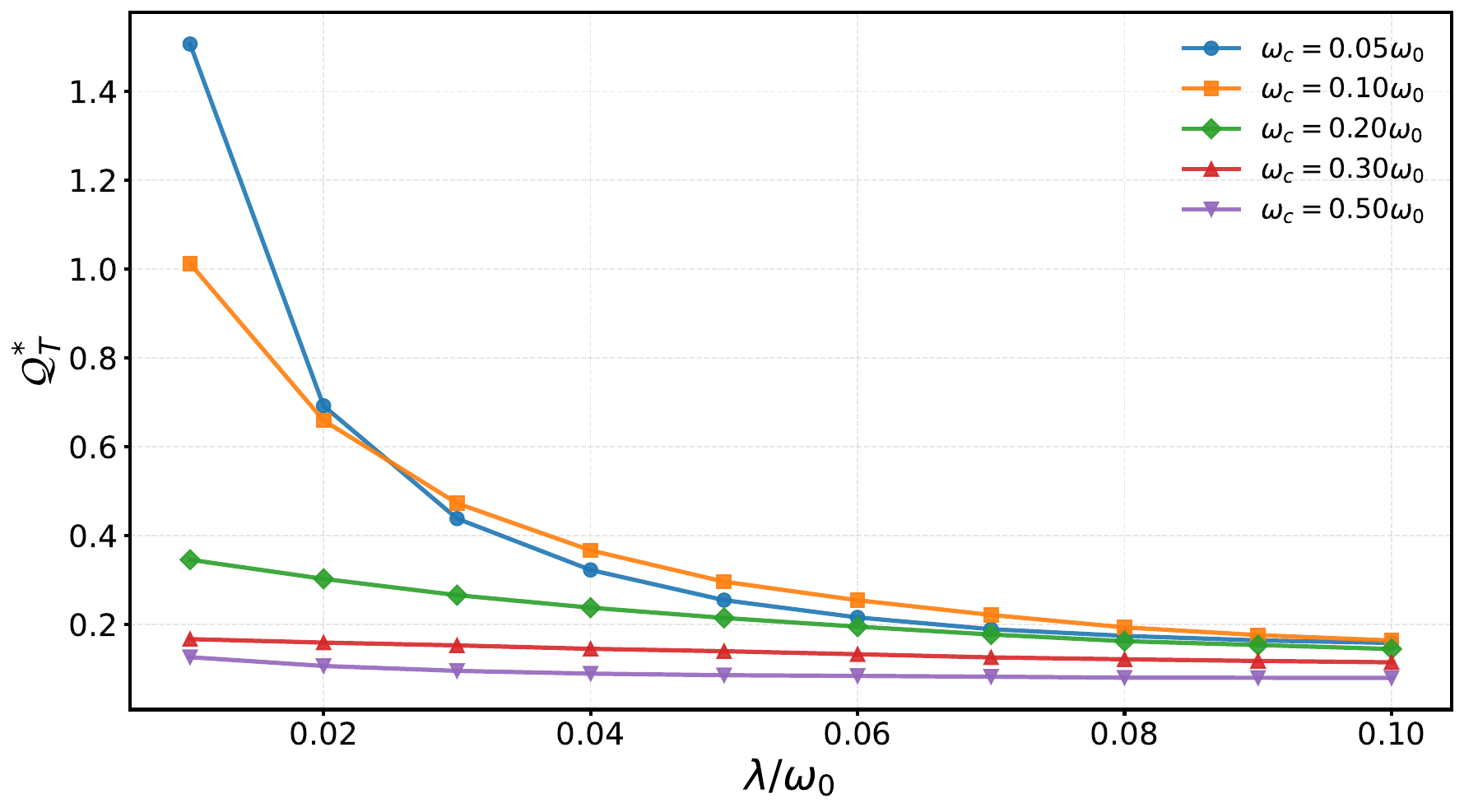}
	\caption{
	The variation of optimal QSNR $\mathcal{Q}^{*}_T$ with respect to $\lambda$ under different $\omega_c$ values, with $T/\omega_0 = 0.2$.
	}\label{fig2}
\end{figure}

    \begin{figure}[bh]
	\centering\includegraphics[width=8.8cm]{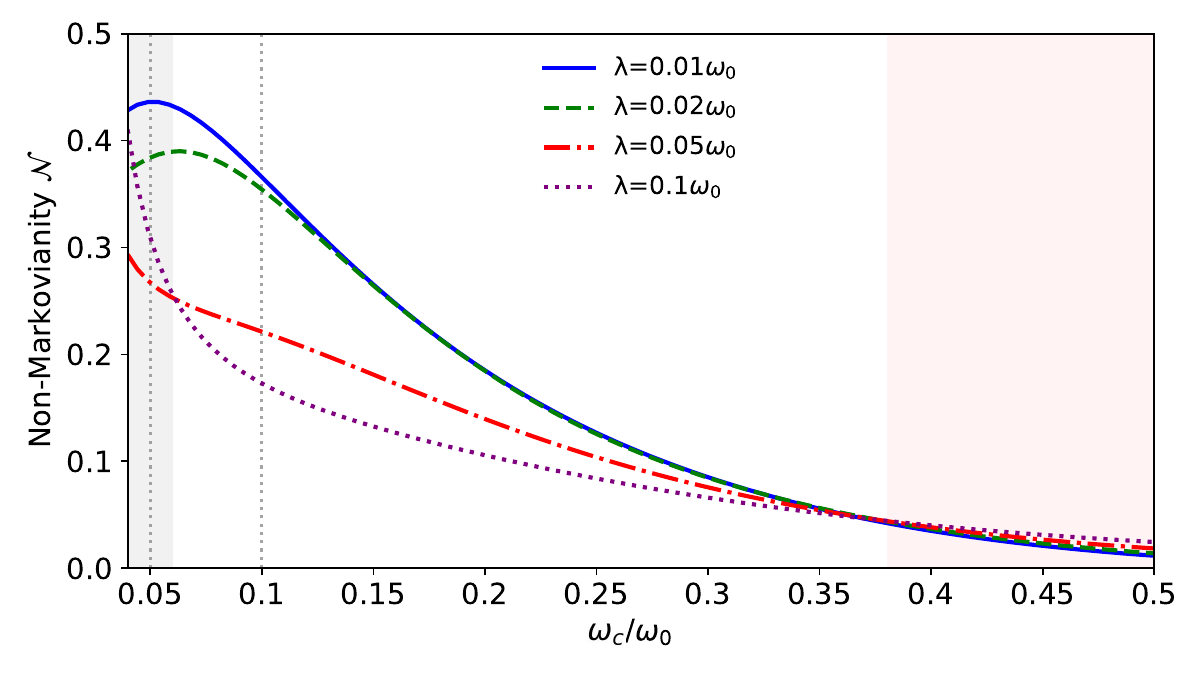}
	\caption{
     Optimal non-Markovianity measure $\mathcal{N}$, evaluated using the orthogonal states $|\pm\rangle = (|0\rangle \pm |1\rangle)/\sqrt{2}$, as a function of $\omega_c$ for varying $\lambda$. The gray-shaded region corresponds to the low cutoff frequency  regime $\omega_c/\omega_0 < 0.06$, while the pink-shaded region denotes high cutoff frequency regime $\omega_c/\omega_0 >0.38$. All other parameters match those in Fig.~\ref{fig1}.
	}\label{fig3}
    \end{figure}

	\section{Optimal coupling strength for non-Markovian quantum thermometry}\label{sec3}

     In this section, we employ the HEOM method to rigorously investigate the impact of different coupling strengths on the QSNR in non-Markovian environments.

	\subsection{ Effect of coupling strength on QSNR dynamics}

     Figure~\ref{fig1} compares the QSNR dynamics obtained via the HEOM method with those predicted by the traditional Bloch-Redfield master equation (BRME), which is based on the Born-Markov approximation~\cite{breuer2002theory,PhysRevA.87.052138} (also see Appendix \ref{appb} for details). These two approaches yield qualitatively different results. 
     In the HEOM simulations, the QSNR displays pronounced oscillatory dynamics in the transient regime before reaching a steady-state value, which is governed by both the system-bath coupling strength and the reservoir’s memory effects—the latter being determined by the cutoff frequency.  
     As the cutoff frequency \(\omega_c\) increases, the system exhibits weaker non-Markovian behavior, and the QSNR dynamics show a monotonic decrease in the optimal value.
    
     In contrast, the BRME predicts a monotonic relaxation toward thermal equilibrium, 
     with a steady-state QSNR that remains independent of the system-bath coupling strength.
     This behavior reflects the intrinsic Markovian assumption of the BRME, which neglects memory effects and thus fails to capture oscillations or transient enhancements in the QSNR. Within this framework, increasing either \(\lambda\) or \(\omega_c\) merely accelerates the approach to equilibrium without altering the final state or introducing non-Markovian features.
     These findings underscore the crucial role of non-Markovian dynamics in enhancing the QSNR. They also highlight the limitations of the BRME in accurately describing quantum thermometric performance in regimes where system-environment correlations and memory effects are significant.

     Figure~\ref{fig2} displays the $\lambda$-dependence of the optimal QSNR $\mathcal{Q}^{*}_T$ for various $\omega_c$ values. A clear trend emerges: for small $\omega_c$ (indicating long environmental memory times), the  $\mathcal{Q}^{*}_T$  shows pronounced sensitivity to weak coupling strengths, a direct manifestation of strong non-Markovianity. However, when $\omega_c$ exceeds $\omega_c/\omega_0=0.3$, the non-Markovian influence weakens considerably, resulting in a nearly flat $\lambda$-dependence of  $\mathcal{Q}^{*}_T$. This behavior culminates in the  $\mathcal{Q}^{*}_T$ approaching its steady-state value, as evidenced in Fig.~\ref{fig1}(c).

     A natural question arises: does stronger environmental non-Markovianity necessarily lead to a higher QSNR? To address this, we evaluate the degree of non-Markovianity \( \mathcal{N} \) using the Breuer–Laine–Piilo (BLP) measure~\cite{PhysRevLett.103.210401}, computed via the  HEOM method (see Appendix~\ref{appc} for computational details). Figure~\ref{fig3} shows the optimized values of \( \mathcal{N} \)  as a function of  spectral cutoff frequency \( \omega_c \) for several fixed values of \( \lambda \). The results reveal three qualitatively distinct regimes. 
     In the low-cutoff-frequency regime (\( \omega_c < 0.06\omega_0 \)), we observe significantly enhanced non-Markovianity \( \mathcal{N} \).  Notably, the dependence of \( \mathcal{N} \) on \(\lambda\) exhibits non-monotonic behavior with characteristic crossover points.
     As the reservoir memory shortens in the intermediate regime (\( 0.06\omega_0 < \omega_c < 0.38\omega_0 \)), the non-Markovianity measure \( \mathcal{N} \) displays an inverse relationship with coupling strength, where weaker coupling paradoxically yields more pronounced non-Markovian character.
     At high-cutoff frequencies (\( \omega_c > 0.38\omega_0 \)),  \( \mathcal{N} \) nearly vanishes for all coupling strengths while exhibiting a weak monotonic increase with growing $\lambda$.
     These systematic observations reveal a sophisticated, nonlinear interplay between system-bath coupling strength and environmental memory timescales in governing quantum non-Markovian dynamics.

     To elucidate the role of non-Markovianity $\mathcal{N}$ in quantum thermometry, we perform a systematic correlation analysis between its $\omega_c$-dependent evolution (Fig.~\ref{fig3}) and the QSNR response (Fig.~\ref{fig1}(a)-(c)).
     In the weak-to-moderate coupling regime (\(\lambda/\omega_0 = 0.01, 0.02, 0.05\)), we find a distinct positive correlation for fixed $(\omega_c/\omega_0= 0.05,0.1)$: larger values of \( \mathcal{N} \) are associated with more pronounced transient peaks in the QSNR.
     This suggests that moderate memory effects can be exploited to enhance the precision of quantum parameter estimation. A similar correlation persists in the intermediate-memory regime (\( 0.06\omega_0 < \omega_c < 0.38\omega_0 \)), where the finite reservoir memory contributes constructively to the thermometric sensitivity.

     However, this trend does not extend uniformly into the strong coupling regime or when the reservoir correlation time becomes short ($\omega_c/\omega_0>0.38$). For example, at \( \lambda = 0.1\omega_0 \) and fixed cutoff frequency \( \omega_c = 0.05\omega_0 \), Fig.~\ref{fig3} still shows a relatively high degree of non-Markovianity. Nevertheless, the corresponding QSNR does not attain its maximum value. For $\omega_c/\omega_0>0.38$,
     we observe an anti-correlation between 
     $\mathcal{N}$ and QSNR.
     Instead, the short-time dynamics display pronounced initial oscillations that rapidly decay to a steady state. This behavior suggests that while strong non-Markovianity can transiently amplify the QSNR, enhanced system-bath coupling accelerates quantum coherence decay, ultimately limiting long-term information retrieval.

     While non-Markovianity can enhance quantum thermometric performance, our analysis reveals a non-monotonic relationship that depends on more than just the magnitude of \( \mathcal{N} \). Importantly, the QSNR demonstrates a stronger correlation with the system-environment coupling strength, where weaker couplings are shown to be more favorable for achieving enhanced temperature sensitivity.
     
\begin{figure}[tbp]
	\centering\includegraphics[width=8.2cm]{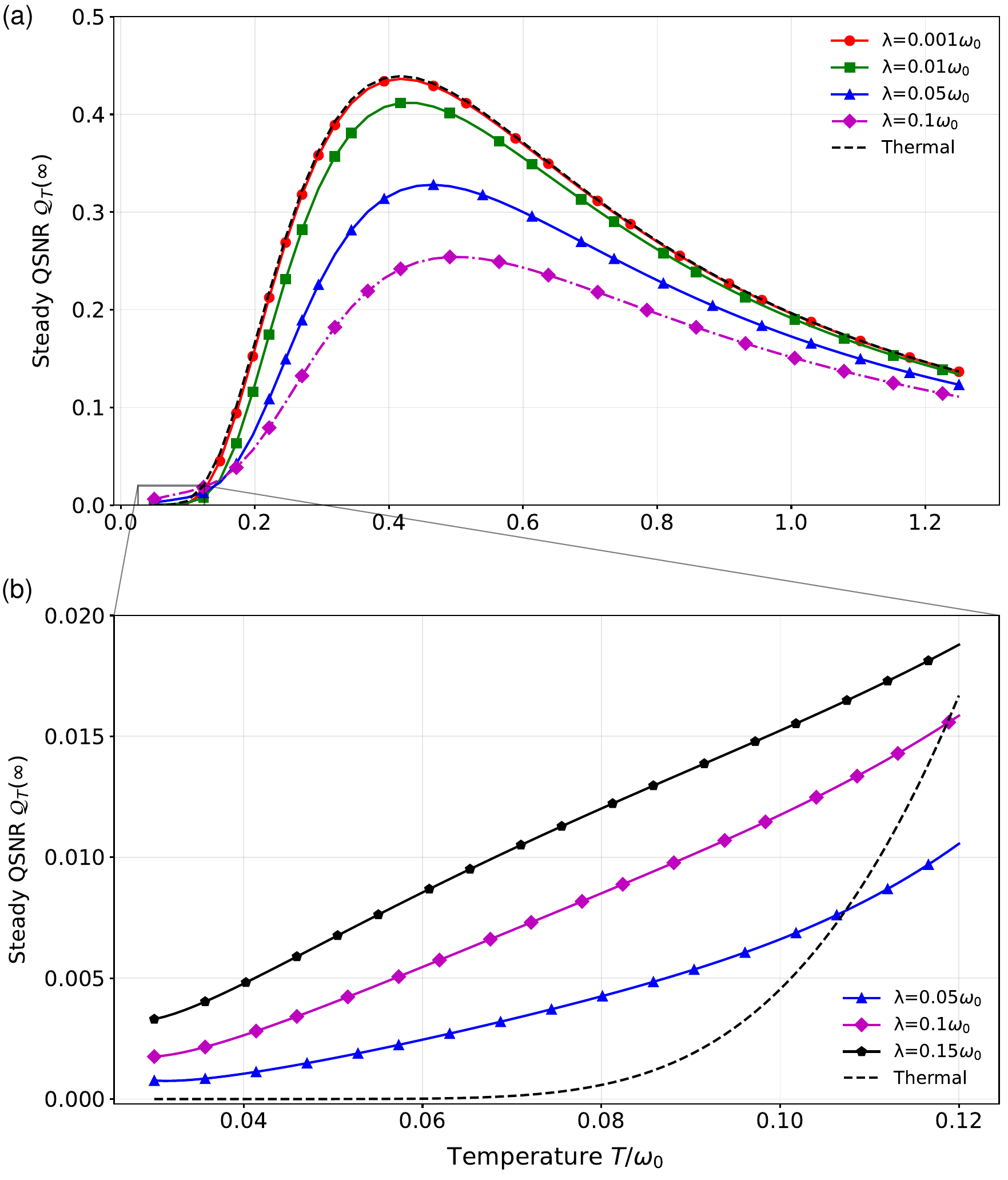}
	\caption{
		(a) Steady-state QSNR \(\mathcal{Q}_T(\infty)\) as a function of temperature \(T/\omega_0\) for varying system-bath coupling strengths \(\lambda/\omega_0\), computed using the HEOM method. Dashed curves indicate thermal equilibrium values for reference, which reveals that complete thermalization occurs only at \(\lambda/\omega_0 \lesssim 0.001\), with stronger couplings exhibiting pronounced deviations from equilibrium. (b) Low-temperature regime (\(T/\omega_0 < 0.12\)), highlighting the emergence of enhanced \(\mathcal{Q}_T(\infty)\) under strong coupling conditions.
	}
	\label{fig4}
\end{figure}
	\subsection{Thermodynamic limits of steady-state QSNR}
	
     The results in Fig.~\ref{fig1} show that although the QSNR reaches a steady-state value in the HEOM simulations, the asymptotic behavior exhibits clear dependence on the coupling strength. To quantitatively assess these values, we first establish a theoretical benchmark using the Bloch-Redfield formalism, which yields the thermal equilibrium limit:  
	 \begin{equation}
      Q^{\mathrm{therm}}_T(\infty) = \left(\frac{\omega_0}{2T}\right)^2 \mathrm{sech}^2 \left(\frac{\omega_0}{2T}\right) \propto C_V(T),
	 \end{equation}
    where \(C_V(T)\) denotes the probe’s heat capacity~\cite{Paris_2016,PhysRevA.108.032220}. This equilibrium QSNR decays exponentially, $Q^{\mathrm{therm}}_T(\infty) =\omega_0^2/(2 T^2) e^{-\omega_0/T} + O(e^{-2\omega_0/T})$,  as \(T/\omega_0 \ll 1\), rendering temperature estimation increasingly challenging in the low-temperature limit.

    Figure~\ref{fig4} presents the steady-state QSNR, \( \mathcal{Q}_T(\infty) \), as a function of the normalized temperature \( T/\omega_0 \) for various values of \( \lambda/\omega_0 \), computed using the HEOM method. The results reveal two qualitatively distinct regimes. In the moderate-to-strong coupling range (\( \lambda > 10^{-3} \omega_0 \)), the QSNR remains suppressed below the thermal bound for \( T \gtrsim 0.15 \omega_0 \) (see Fig.~\ref{fig4}(a)). These results demonstrate that weak system-environment coupling (such as \(\lambda =10^{-3}\omega_0\)) is thermodynamically favorable in the intermediate-temperature regime, as it both facilitates equilibration and enhances the precision of steady-state quantum thermometry. Moreover, we observe that when the system-environment coupling remains sufficiently weak (\(\lambda \leq 10^{-3}\omega_0\)), the steady-state solutions obtained from both methods exhibit remarkable consistency. Specifically, the BRME steady-state solution demonstrates complete agreement with the corresponding HEOM result under these conditions.

    In sharp contrast, when the temperature drops below \( T < 0.1 \omega_0 \), and the system–bath coupling becomes strong (\( \lambda > 0.05 \omega_0 \)), the system exhibits a clear departure from thermal equilibrium behavior. As shown in Fig.~\ref{fig4}(b), the QSNR demonstrates a power-law scaling with temperature, as also reported in Ref.~\cite{PhysRevA.96.062103}. For sufficiently strong coupling (\( \lambda > 0.1 \omega_0 \)), this scaling becomes nearly linear in our model. Moreover, the magnitude of the QSNR increases monotonically with the coupling strength, violating the exponential suppression that typically characterizes equilibrium thermometry at low temperatures. 
    This anomalous behavior suggests the existence of non-canonical system–environment correlations that sustain metrological sensitivity beyond thermal equilibrium limits. These findings align with theoretical predictions from the reaction coordinate formalism~\cite{PhysRevA.108.032220} and polaron-based analyses~\cite{PhysRevA.111.052623}, both of which show that strong system–bath coupling can enhance thermometric precision at ultra-low temperatures. Crucially, these frameworks highlight how pronounced quantum correlations emerge between the system and its environment in the strong coupling regime. Our results thus demonstrate that, for steady-state quantum thermometry in the ultra-low temperature regime, strong coupling can be exploited to generate non-equilibrium quantum correlations—transforming them into a metrological resource that surpasses the fundamental sensitivity bounds of thermal equilibrium.

    Building on our analysis of both the dynamical and steady-state behavior of quantum thermometry, our findings lead to the following conclusion: while non-Markovianity can enhance dynamical sensitivity, strong system–bath coupling does not necessarily induce stronger memory effects. The relationship between non-Markovianity and the QSNR is more intricate than a simple proportional dependence. In non-Markovian environments, weak coupling is generally more advantageous for dynamical sensing, as it gives rise to pronounced transient features that enhance estimation precision. In contrast, strong coupling proves beneficial primarily in the ultra-low-temperature steady-state regime, where it induces non-canonical system–environment correlations that enable sensitivity beyond the limitations imposed by thermal equilibrium.

\section{Quantum Particle Swarm Optimization for Maximizing QSNR}
\label{sec4}
       \begin{table*}[htbp]
  	  \centering
	   \caption{Comparison of key components between PSO and QPSO.}
	  \begin{tabular}{p{2.8cm} | p{5.8cm} | p{6cm}}
		\hline
		\textbf{Concept} & \textbf{PSO} & \textbf{QPSO} \\
		\hline
		\textbf{Particle} $\mathbf{X}_i$ & Represents a candidate solution in a $D$-dimensional parameter space, e.g., control amplitudes or pulse timings. & Same interpretation, but particle position is updated via a probabilistic (quantum-inspired) rule rather than classical velocity. \\
		\hline
		\textbf{Velocity} $\mathbf{V}_i$ & Governs classical particle motion; position is updated as $\mathbf{X}_i^{t+1} = \mathbf{X}_i^t + \mathbf{V}_i^t$. & No explicit velocity; motion arises from quantum tunneling behavior simulated through stochastic sampling. \\
		\hline
		\textbf{Personal best} $\mathbf{P}_i$ & Best position found so far by particle $i$ based on fitness value. & Same as in PSO. \\
		\hline
		\textbf{Global best} $\mathbf{G}$ & Best overall position found by all particles. & Same as in PSO. \\
		\hline
		\textbf{Fitness function} $f(\mathbf{X})$ & Quantifies solution quality, e.g., QSNR or QFI at $\mathbf{X}_i$. Guides the optimization. & Same function. Used identically to evaluate candidate solutions. \\
		\hline
		\textbf{Position update rule} & Deterministic: Eq.~(\ref{eq:pso_update}). & Probabilistic: Eq.~(\ref{qxl}). \\
		\hline
		\textbf{Exploration mechanism} & Particles exchange information via local/global best and adjust velocities accordingly. & Particles probabilistically collapse toward mean best position, allowing broader search and escape from local optima. \\
		\hline
	\end{tabular}
	\label{tab:PSO_QPSO_compare}
\end{table*}

     To further explore the effects of coupling beyond the weak regime, we will develop an optimized hybrid framework that integrates the HEOM method with QPSO to maximize the QSNR. This approach aims to exploit the tunability of the probe qubit’s energy-level structure to achieve optimal sensitivity in nonequilibrium quantum thermometry.

     Particle Swarm Optimization (PSO) is a widely used bio-inspired algorithm that models the collective behavior observed in natural systems such as bird flocks and fish schools~\cite{Wang2018,PhysRevLett.104.063603,PhysRevA.110.023718}. In PSO, a population of agents—referred to as particles—collaboratively explores a $D$-dimensional solution space to locate the global optimum of a given objective (fitness) function.
     Each particle $i$ at iteration $t$ is characterized by a position vector $\mathbf{X}_i(t) = (X_{i1}, \dots, X_{iD})$, representing a candidate solution, and a velocity vector $\mathbf{V}_i(t) = (V_{i1}, \dots, V_{iD})$, which determines its direction and speed of motion. Additionally, each particle retains a personal best position $\mathbf{P}_i(t)$, corresponding to the best solution it has encountered so far, while the swarm collectively maintains a global best position $\mathbf{G}(t)$, representing the best solution found across all particles.
     The position and velocity of each particle are updated according to:
    \begin{subequations}\label{eq:pso_update}
	\begin{align}
		\mathbf{X}_i^{t+1} &= \mathbf{X}_i^t + \mathbf{V}_i^{t+1}, \label{eq:position_update} \\
		\mathbf{V}_i^{t+1} &= {\rm w}\mathbf{V}_i^t + f_1 r_1 (\mathbf{P}_i^t - \mathbf{X}_i^t) + f_2 r_2 (\mathbf{G}^t - \mathbf{X}_i^t). \label{eq:velocity_update}
	\end{align}	
   \end{subequations}
    where ${\rm w}$ is the inertia weight that balances exploration and exploitation; $f_1$ and $f_2$ are cognitive and social learning coefficients, respectively; and $r_1, r_2 \in [0, 1]$ are independent random variables introducing stochasticity into the search dynamics. The fitness function evaluates the quality of each candidate solution and guides the swarm toward optimal regions in the parameter space.

     QPSO generalizes classical PSO by introducing a quantum-inspired probabilistic update mechanism~\cite{1330875}. Unlike classical PSO, QPSO does not explicitly retain a velocity vector \(\mathbf{V}_i\) of each particle $i$; instead, particle movement is governed by quantum transition probabilities derived from an effective potential well.
     Each particle is modeled as a quantum entity confined in a delta-function potential well centered at its personal best position \(\mathbf{P}_i \). The associated ground-state wavefunction is given by \(\Psi(\mathbf{X}_{i}) = \frac{1}{\sqrt{L}} e^{ -|\mathbf{X}_{i} - \mathbf{P}_{i}|/L }\), leading to a probability density \(Q(\mathbf{X}_{i}) = \frac{1}{L} e^{ -2|\mathbf{X}_{i} - \mathbf{P}_{i}|/L }\), where \(L\) characterizes the spatial extent of the wavepacket. Particle positions are updated by sampling from this distribution, simulating a quantum measurement process. 
     The position update rule is given by~\cite{1330875}:
     \begin{equation}\label{qxl}
     	\mathbf{X}_{i}^{t+1} = \mathbf{P}_{i}^t \pm \alpha \left| \mathbf{mbest} - \mathbf{X}_{i}^t \right| \ln(1/u),
     \end{equation}
     where $\mathbf{mbest} = \frac{1}{M} \sum_{i=1}^{M} \mathbf{P}_{i}$ denotes the mean best position defined as the average of all personal best positions in the population, $M$ is the swarm size (number of particles), $u \sim \mathcal{U}(0,1]$ is a uniformly distributed random variable, and $\alpha \in [0,1]$ is the contraction-expansion coefficient that controls the trade-off between global exploration and local exploitation.
     
     This formulation enables particles to exhibit tunneling-like behavior and long-range transitions, enhancing the algorithm’s ability to escape local optima and improving global convergence. As a result, QPSO has demonstrated superior performance compared to classical PSO in a wide range of complex optimization tasks.
     Table 1 presents a comparative analysis of the key concepts and exploration mechanisms between PSO and QPSO.

\begin{figure*}[tp]
	\centering\includegraphics[width=17.6cm]{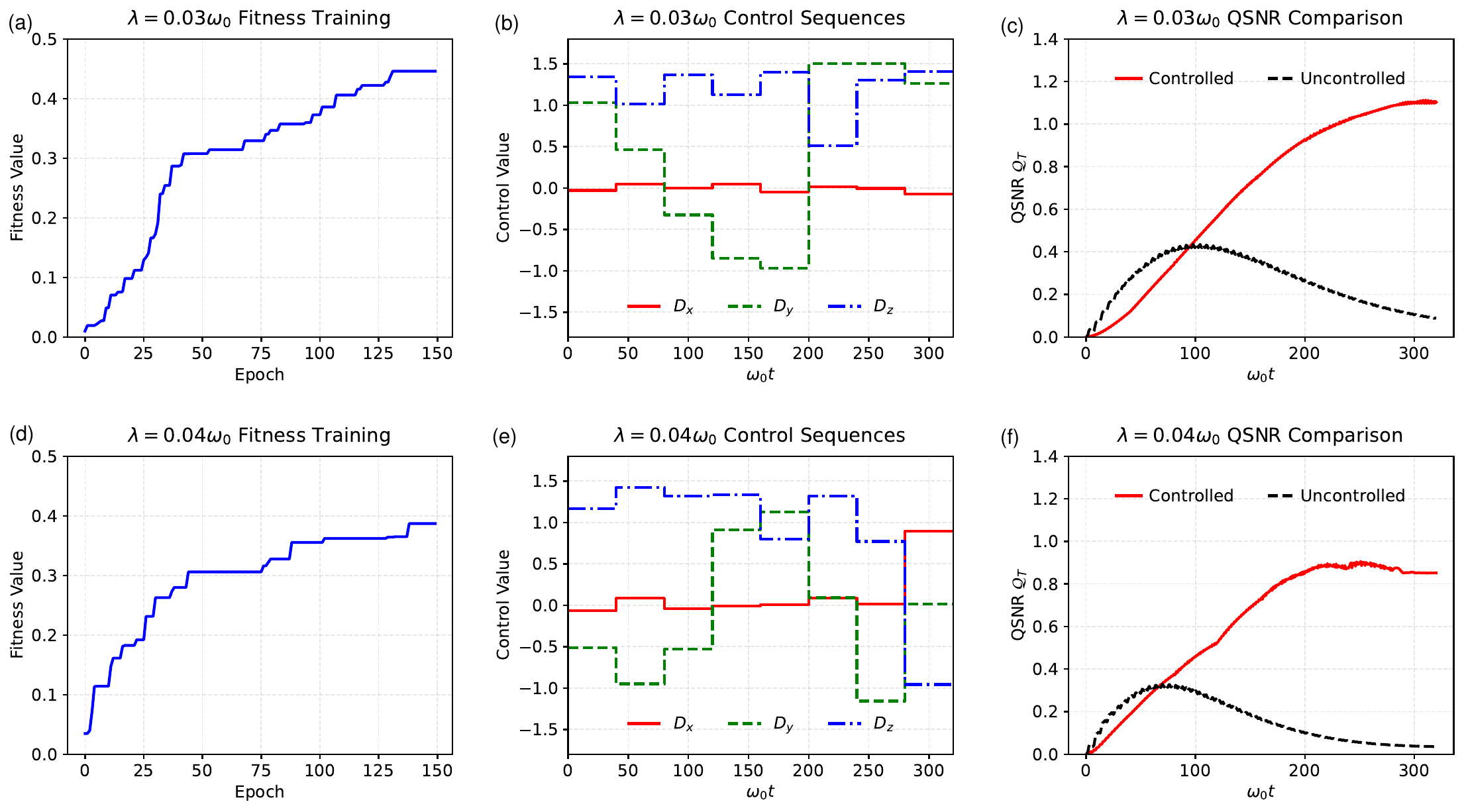}
	\caption{ 
		(a-c) Optimization of QSNR using the QPSO algorithm  for  \(\lambda = 0.03\omega_0\): (a) Evolution of the fitness value over optimization epochs; (b) Control parameter sequence corresponding to the optimal fitness value; (c) Comparison of the QSNR dynamics with and without optimal control.
		(d-f) Optimization of QSNR using the QPSO algorithm for \(\lambda = 0.04\omega_0\): (d) Evolution of the fitness value over optimization epochs; (e) Control parameter sequence corresponding to the optimal fitness value; (f) Comparison of the QSNR dynamics with and without optimal control. Parameters: $\omega_c=0.05\omega_0$ and $T/\omega_0=0.2$.
	}
	\label{fig5}
\end{figure*}

     In our quantum thermometry optimization scheme, QPSO offers an efficient strategy for optimizing time-dependent Hamiltonians governed by piecewise-constant control fields. As a representative example, we consider a single-qubit system driven by \(N\) time-segmented control fields, described by the Hamiltonian~\cite{Xu2019,PhysRevA.96.012117,Xiao2022}
	\begin{equation}\label{ham}
	\hat{H}_S(t) = \frac{\omega_0}{2}\hat{\sigma}_z + \frac{1}{2} \sum_{k=1}^N    \left[\sum_{i=x,y,z} D_i^{(k)}(t)\hat{\sigma}_i  \right],
\end{equation}
     where the real-valued parameters \(\{D_x^{(k)}, D_y^{(k)}, D_z^{(k)}\}\) define the control amplitudes within each segment. To accurately capture the system’s non-Markovian dynamics and its interaction with the environment, we employ the HEOM formalism as the underlying simulation framework.

     The QPSO algorithm is then tasked with optimizing the control parameters by treating them as coordinates in a \(3N\)-dimensional search space. Each particle encodes a complete control sequence as \(\mathbf{X}_i = (D_x^{(1)}, D_y^{(1)}, D_z^{(1)}, \ldots, D_x^{(N)}, D_y^{(N)}, D_z^{(N)})\). The fitness function is defined by a time-weighted average of the QSNR \(\mathcal{Q}_T(t)\) over the evolution interval \(t \in [0, t_{\mathrm{max}}]\),
     \begin{equation}
	   \mathcal{Q}_T^{\mathrm{weighted}} = \frac{1}{N_t} \sum_{i=0}^{N_t-1} w_j \mathcal{Q}_T(t),
     \end{equation}
     where \(w_j = (j+1)/N_t\) introduces a linear weighting scheme that prioritizes the later stages of evolution, while maintaining smooth QSNR dynamics throughout the entire process.  This design ensures that the optimization process emphasizes long-time coherence and measurement sensitivity, which are essential for robust quantum control and metrology in realistic noisy environments.
     The QSNR is computed directly from Eq.~(\ref{qsnr}).  This indicates that, in practice, achieving a temperature sensing precision significantly higher than that of steady-state thermometers via dynamic optimization necessarily requires consuming more measurement resources. Here, we do not address the design of the optimal measurement scheme itself, but rather emphasize the potential of optimization to reach the ultimate theoretical limits.

     Through the integration of quantum-inspired global optimization and accurate open-system dynamics, the QPSO-HEOM framework offers a practical and scalable approach for high-fidelity quantum control in the strong coupling, and non-Markovian memory effects.
  	 In the quantum intelligent optimization process shown in Fig.~\ref{fig5}, we employ QPSO to enhance the QSNR under moderate coupling strength. Specifically, the total evolution time \( t = 320/\omega_0 \) is divided into $N=8$ equal segments. This leads to an optimization problem in a 24-dimensional control space (3 parameters per segment over 8 segments). The QPSO algorithm operates with a swarm size of 20 particles, and the optimization is carried out over 150 iterations [see Figs.~\ref{fig5}(a) and \ref{fig5}(d)].

	 As illustrated in Fig.~\ref{fig5}, we investigate optimal control protocols for two distinct coupling strengths, \( \lambda = 0.03\omega_0 \) and \( \lambda = 0.04\omega_0 \). For each case, the QPSO algorithm converges to an optimal sequence of control  parameters \( \{D_x^{(k)}, D_y^{(k)}, D_z^{(k)}\} \) [Figs.~\ref{fig5}(b) and \ref{fig5}(e)], which yields a significant enhancement of the QSNR compared to the uncontrolled dynamics.  Remarkably, these improvements remain robust in the long-time limit, indicating the stability of the optimized control sequences. A comparative analysis reveals more pronounced QSNR enhancement for the weaker coupling case (\( \lambda = 0.03\omega_0 \)), suggesting greater controllability in this regime.

     Compared to reinforcement learning (RL)—a widely used, versatile machine learning framework based on autonomous agent-environment interactions~\cite{PhysRevX.8.031084,PhysRevLett.126.060401,Ai2022,Zhang2023,PhysRevA.109.042417,PhysRevA.111.063504,PhysRevA.111.053517}—our method provides some advantages  for the particular problem at hand.
     While RL typically operates within the Markov decision process (MDP) framework, recent research has explored methods for handling non-Markovian environments. One approach uses reward machines, guided by deterministic finite automata, to transform non-Markovian reward functions into a learnable Markovian form~\cite{NIPS1990_70c639df}. This method addresses history-dependent rewards and accelerates learning with human input, while maintaining convergence guarantees. Other studies use specialized architectures like recurrent neural networks (RNNs) to capture temporal dependencies~\cite{Mizutani2017,Banchi_2018}. In contrast, our optimization scheme is particularly efficient for problems with limited control sequences, achieving near-optimal solutions within just 150 iterations, demonstrating faster convergence than conventional RL for this problem class.

	\section{conclusions and outlook}\label{sec5}

     In conclusion, we have presented a comprehensive study of quantum thermometry in non-Markovian environments using a single-qubit probe. Through numerically exact HEOM simulations, we show that strong system–environment coupling does not necessarily enhance thermometric precision in our proposed model. Our results reveal that the relationship between non-Markovianity and the QSNR is more intricate than a simple proportional dependence, with weak coupling proving particularly advantageous for dynamical temperature sensing. In contrast, strong coupling primarily benefits steady-state thermometry at ultralow temperatures by mitigating the limitations of thermal equilibrium. As the temperature approaches zero, the QSNR shifts from an exponential to an approximately linear decay, substantially extending the operational range of quantum thermometry into the ultralow-temperature regime.
	
     To optimize the QSNR in non-weak coupling regimes, we developed a hybrid optimization framework that combines HEOM dynamics with quantum-inspired particle swarm optimization. This approach allows for systematic tuning of the effective probe-environment interaction, capturing the complex interplay between coupling strength and environmental memory effects. The framework proves to be a powerful tool for designing optimized quantum thermometers in realistic non-Markovian environments.
     
     These results significantly advance our understanding of how coupling strength and environmental memory can be harnessed for quantum metrology. The insights obtained offer valuable design principles for quantum thermometry, particularly in identifying optimal coupling regimes for various sensing tasks.

     Importantly, the HEOM framework employed here is equally applicable to multi-qubit thermometric probes. As illustrated in Appendix~\ref{appd}, our preliminary analysis of a two-qubit system reveals that weak coupling similarly enhances the dynamical QSNR, and that the associated behavior exhibits richer dynamics compared to the single-qubit case. The accuracy of non-Markovian quantum thermometry is influenced by both the form of interaction between the probe qubits and the coupling between the system and the environment.
     Future research will aim to extend these principles to multi-qubit probes, structured reservoirs, and finite-time metrological protocols, where memory effects may provide additional advantages.

\appendix

\begin{widetext}
	\section{Derivation of the HEOM for a Drude-Lorentz spectral at finite temperature} \label{appa}
	
	We derive the HEOM for a system coupled to a bosonic environment with a Drude-Lorentz spectral density at finite temperature.
	The total Hamiltonian is
	\begin{equation}
		\hat{H} = \hat{H}_S + \hat{H}_{SB} + \hat{H}_B,
	\end{equation}
	where
	\[
	\hat{H}_B = \sum_k \omega_k \hat{a}_k^\dagger \hat{a}_k, \quad
	\hat{H}_{SB} = \hat{S} \otimes \hat{B}, \quad
	\hat{B} = \sum_k ( g_k \hat{a}_k^\dagger + g_k^* \hat{a}_k ).
	\]
	
	In the interaction picture with respect to \( \hat{H}_S + \hat{H}_B \), the evolution of the total density matrix is given by
	\begin{equation}
		\frac{d}{dt} \hat{\rho}_{\text{tot}}(t) = -i \hat{H}_{SB}^{\times}(t)\hat{\rho}_{\text{tot}}(t), \quad
		\hat{\rho}_{\text{tot}}(t) = \mathcal{T} \exp\left( -i \int_0^t d\tau\, \hat{H}_{SB}^{\times}(\tau) \right) \hat{\rho}_{\text{tot}}(0),
	\end{equation}
	where \( \mathcal{T} \) denotes the time-ordering operator.
	
	Assuming an initially factorized state \( \hat{\rho}_{\text{tot}}(0) =\hat{\rho}_S(0) \otimes \hat{\rho}_B \), the reduced system density matrix is
	\begin{equation}
		\hat{\rho}_S(t) = \mathrm{Tr}_B \left\{ \mathcal{T} \exp\left( -i \int_0^t d\tau\, \hat{H}_{SB}^{\times}(\tau) \right) \hat{\rho}_S(0) \otimes \hat{\rho}_B \right\}.
	\end{equation}
	
	For a Drude-Lorentz spectral density, the reservoir correlation function can be expanded as
	\begin{equation}
		C(t) = \langle \hat{B}(t) \hat{B}(0) \rangle_B = \sum_{k=0}^{\infty} c_k e^{-\nu_k t},
	\end{equation}
	where \( \nu_k \) are Matsubara frequencies, and \( c_k \in \mathbb{C} \) are expansion coefficients.
	
	This exponential structure allows us to recast the reduced dynamics as a time-ordered exponential:
	\begin{equation}
		\hat{\rho}_S(t) = \mathcal{T}\exp\left\{-\int_0^t d\tau_2\, \hat{S}^{\times}(\tau_2) \left[ \int_0^{\tau_2} d\tau_1\, s(\tau_2 - \tau_1) \hat{S}^{\times}(\tau_1) - \frac{i}{2} \chi(\tau_2 - \tau_1) \hat{S}^{\circ}(\tau_1) \right] \right\} \hat{\rho}_S(0),
	\end{equation}
	where the commutator and anti-commutator superoperators are defined by
	\[
	\hat{S}^{\times}(t)\hat{O} = [\hat{S}(t), \hat{O}], \quad
	\hat{S}^{\circ}(t)\hat{O} = \{ \hat{S}(t), \hat{O} \},
	\]
	and the real and imaginary parts of the correlation function are
	\begin{equation}
		s(t) = \sum_{k=0}^\infty \text{Re}(c_k) e^{-\nu_k t}, \quad
		\chi(t) = -2 \sum_{k=0}^\infty \text{Im}(c_k) e^{-\nu_k t}, \quad t \geq 0.
	\end{equation}
	
	We introduce the superoperators
	\begin{equation}
		\hat{\Phi}(t) = -i \hat{S}^{\times}(t), \quad
		\hat{\Theta}_k(t) = -i \left[ \text{Re}(c_k) \hat{S}^{\times}(t) + i\, \text{Im}(c_k) \hat{S}^{\circ}(t) \right],
	\end{equation}
	and rewrite the reduced density matrix in the Schr\"odinger picture as
	\begin{equation}
		\hat{\rho}_S(t) = \hat{U}(t)\, \mathcal{T} \exp \left( \int_0^t d\tau\, \hat{\Phi}(\tau) \sum_k \int_0^\tau d\tau'\, e^{-\nu_k (\tau - \tau')} \hat{\Theta}_k(\tau') \right) \hat{\rho}_S(0)\, \hat{U}^\dagger(t),
	\end{equation}
	with \( \hat{U}(t) = e^{-i(\hat{H}_S + \hat{H}_B)t} \).
	
	Differentiating with respect to time yields the first-level equation:
	\begin{equation}
		\frac{d}{dt}\hat{ \rho}_S(t) = -i \hat{H}_S^{\times} \rho_S(t) + \hat{\Phi}(t) \sum_k \hat{\varrho}_{\vec{e}_k}(t),
	\end{equation}
	where \( \hat{\varrho}_{\vec{e}_k}(t) \) are the auxiliary density operators (ADOs), defined as
	\begin{equation}
		\hat{\varrho}_{\vec{e}_k}(t) = \hat{U}(t)\, \mathcal{T} \left\{ \int_0^t d\tau\, e^{-\nu_k(t - \tau)} \hat{\Theta}_k(\tau)\, \exp\left( \int_0^t d\tau_2 \int_0^{\tau_2} d\tau_1\, \hat{\Phi}(\tau_2) \sum_m e^{-\nu_m(\tau_2 - \tau_1)}\hat{ \Theta}_m(\tau_1) \right) \right\} \hat{\rho}_S(0)\, \hat{U}^\dagger(t).
	\end{equation}
	
	By differentiating the ADOs recursively, we obtain the HEOM:
	\begin{align}
		\frac{d}{dt} \hat{\varrho}_{\vec{e}_k}(t) &= \left( -i \hat{H}_S^{\times} - \nu_k \right) \hat{\varrho}_{\vec{e}_k}(t) +\hat{\Theta}_k(t) \rho_S(t) + \hat{\Phi}(t) \sum_n \hat{\varrho}_{\vec{e}_k + \vec{e}_n}(t), \\
		\frac{d}{dt} \hat{\varrho}_{\vec{e}_k + \vec{e}_n}(t) &= \left( -i \hat{H}_S^{\times} - \nu_k - \nu_n \right) \hat{\varrho}_{\vec{e}_k + \vec{e}_n}(t) + \hat{\Theta}_k(t) \hat{\varrho}_{\vec{e}_n}(t) + \hat{\Theta}_n(t) \hat{\varrho}_{\vec{e}_k}(t) + \hat{\Phi}(t) \sum_m \hat{\varrho}_{\vec{e}_k + \vec{e}_n + \vec{e}_m}(t).
	\end{align}
	The unit vector $\vec{e}_k$ is defined as
	$
	\vec{e}_k= (0,0,\dots,1_k,\dots,0),
	$
	i.e., only the $k$th position is 1, and others are zero.
	
	Repeat the steps used for Eqs. (A11) and (A12), the general form of the HEOM is thus
	\begin{equation}
		\frac{d}{dt} \hat{\varrho}_{\vec{n}}(t)
		= \left( -i \hat{H}_S^{\times} - \sum_k n_k \nu_k \right) \hat{\varrho}_{\vec{n}}(t)
		+ \sum_k\hat{ \Phi}(t) \hat{\varrho}_{\vec{n} + \vec{e}_k}(t)
		+ \sum_k n_k \nu_k \hat{\Theta}_k(t) \hat{\varrho}_{\vec{n} - \vec{e}_k}(t).
	\end{equation}
	This recursive structure provides a non-perturbative, non-Markovian framework for simulating open quantum systems under structured environmental influence.
\end{widetext}

	\section{Derivation of the Bloch-Redfield master equation} \label{appb}

	The Bloch-Redfield master equation describes the dynamics of a quantum system weakly coupled to a thermal environment. This appendix provides a concise derivation, emphasizing key assumptions and results.
		
	Consider a system \( S \) and bath \( B \) with total Hamiltonian:
	\begin{equation}
		\hat{H} =\hat{H}_S + \hat{H}_B + \hat{H}_I, \quad \hat{H}_I = \sum_\alpha \hat{S}_\alpha \otimes \hat{B}_\alpha,
	\end{equation}
	where \( \hat{H}_S \) and \( \hat{H}_B \) are the system and reservoir Hamiltonians, and \( \hat{H}_I \) couples system operators \( \hat{S}_\alpha \) to bath operators \( \hat{B}_\alpha \). The reduced system density matrix is \( \hat{\rho}_S(t) = \text{Tr}_B[\hat{\rho}(t)] \).
		
	In the interaction picture, operators evolve as \( \hat{O}(t) = e^{i(\hat{H}_S + \hat{H}_B)t} \hat{O}e^{-i(\hat{H}_S + \hat{H}_B)t} \), and the interaction Hamiltonian is:
	\begin{equation}
		\hat{H}_I(t) = \sum_\alpha \hat{S}_\alpha(t) \otimes \hat{B}_\alpha(t).
	\end{equation}
	The von Neumann equation \( \dot{\hat{\rho}}(t) = -i [\hat{H}_I(t), \hat{\rho}(t)] \) is solved perturbatively to second order, yielding:
	\begin{equation}
		\frac{d\hat{\rho}_S(t)}{dt} = -\int_0^t ds \, \text{Tr}_B [\hat{H}^{\times}_I(t) \hat{H}^{\times}_I(s) \hat{\rho}(s)].
	\end{equation}
		
	We assume: (1) an initial separable state \( \hat{\rho}(0) =\hat{ \rho}_S(0) \otimes \hat{\rho}_B \), with \( \hat{\rho}_B = e^{-\beta \hat{H}_B}/Z_B \) and \( \text{Tr}_B[\hat{H}_I(t) \hat{\rho}_B] = 0 \); (2) the Born approximation, \( \hat{\rho}(s) \approx \hat{\rho}_S(s) \otimes \hat{ \rho}_B \); and (3) the Markov approximation, replacing \( \hat{\rho}_S(s) \to \hat{\rho}_S(t) \) and extending the integral to \( t \to \infty \).
		
	The reservoir’s influence is captured by correlation functions:
		\begin{equation}
		C_{\alpha\beta}(\tau) = \text{Tr}_B \left[ \hat{B}_\alpha(\tau) \hat{B}_\beta(0) \hat{\rho}_B \right],
		\end{equation}
	with the dissipation rates in the frequency domain:
		\begin{equation}
		\Gamma_{\alpha\beta}(\omega) = \int_0^\infty d\tau \, e^{i \omega \tau} C_{\alpha\beta}(\tau).
	    \end{equation}
	For a thermal reservoir:
	\begin{eqnarray}
		&&\text{Re}[\Gamma_{\alpha\beta}(\omega)] = \pi J_{\alpha\beta}(\omega) \left( 1 + n_\beta(\omega) \right),\\
		 && \text{Im}[\Gamma_{\alpha\beta}(\omega)] = \mathcal{P} \int_{-\infty}^\infty \frac{J_{\alpha\beta}(\omega')}{\omega - \omega'} d\omega',
	\end{eqnarray}
	where \( n_\beta(\omega) = (e^{\beta \omega} - 1)^{-1} \).
		
    Decomposing $\hat{S}_\alpha(t) = \sum_\omega \hat{S}_\alpha(\omega) e^{i\omega t}$, with $\hat{S}_\alpha(\omega) = \sum_{E_m - E_n = \omega} |m\rangle \langle m| \hat{S}_\alpha |n\rangle \langle n|$, and without applying the secular approximation, the Redfield master equation in the Schr\"odinger picture reads:
    \begin{align} \label{rm}
	 \frac{d\hat{\rho}_S}{dt} 
	 &= -i(\hat{H}^{\times}_S + \hat{H}^{\times}_{\mathrm{LS}}) \hat{\rho}_S \nonumber \\
	 &\quad + \sum_{\omega, \omega',\alpha, \beta} \Gamma_{\alpha\beta}(\omega) 
	 \big( \hat{S}_\beta(\omega') \hat{\rho}_S \hat{S}_\alpha^\dagger(\omega) \nonumber \\
	 &\qquad\qquad - \hat{S}_\alpha^\dagger(\omega) \hat{S}_\beta(\omega') \hat{\rho}_S \big) + \text{h.c.}
    \end{align}
   where the Lamb shift Hamiltonian is:
   \begin{equation}
	\hat{H}_{\mathrm{LS}} = \sum_{\omega, \alpha, \beta}     \mathrm{Im}[\Gamma_{\alpha\beta}(\omega)] \hat{S}_\alpha^\dagger(\omega) \hat{S}_\beta(\omega).
   \end{equation}
  Equation~(\ref{rm}) represents the Bloch-Redfield master equation, which
retains off-resonant terms ($\omega \neq \omega'$), and thus captures coherences between different energy eigenstates. It is accurate in the weak-coupling and Markovian regime but may not preserve complete positivity, unlike Lindblad forms obtained under the secular approximation.

\section{Non-Markovianity} \label{appc}
\begin{figure}[th]
	\centering\includegraphics[width=8.6cm]{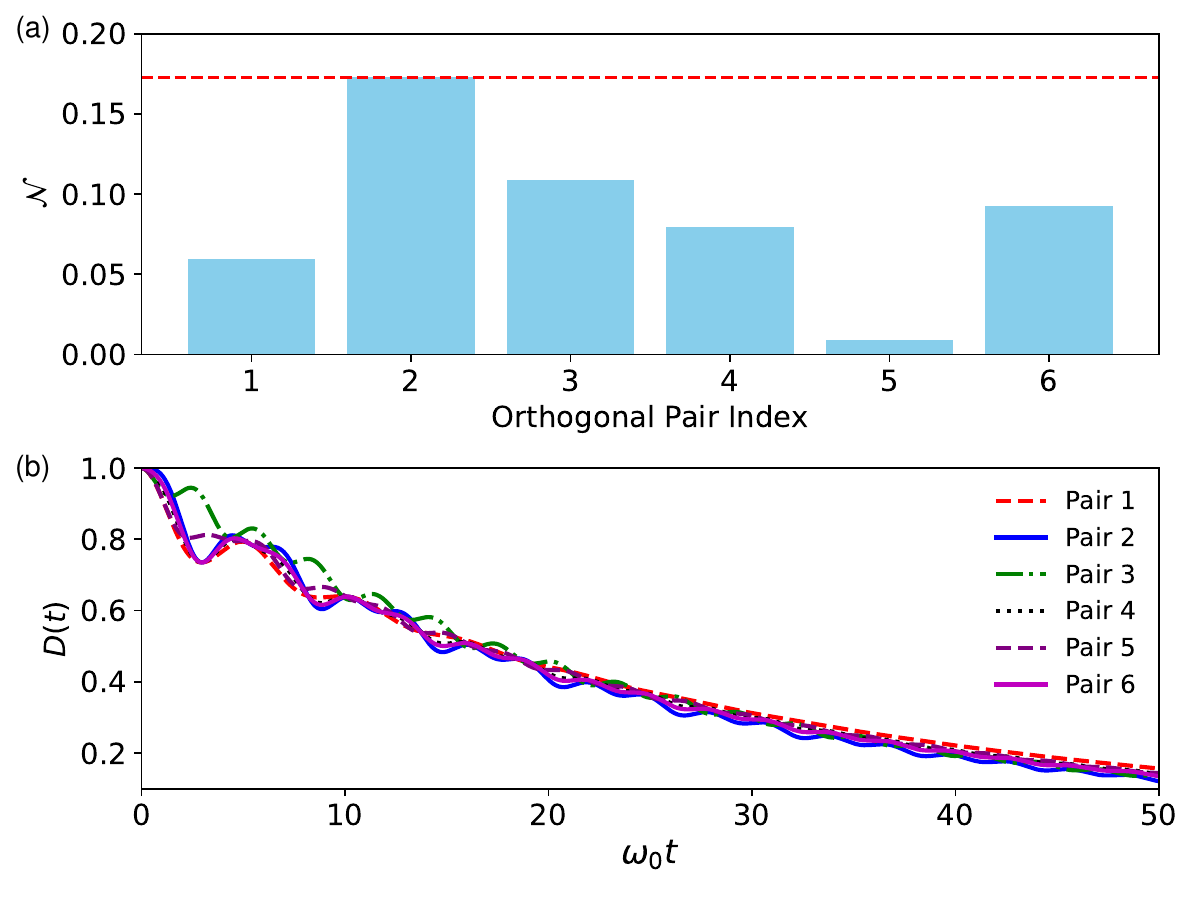}
	\caption{
		(a) Six pairs of orthogonal quantum states are used to probe non-Markovian dynamics: (1) computational basis states $|0\rangle$ and $ |1\rangle$; (2) $x$-axis superpositions $|\pm\rangle= (|0\rangle \pm |1\rangle)/\sqrt{2}$; (3) $y$-axis superpositions $|L,R\rangle = (|0\rangle \pm i|1\rangle)/\sqrt{2}$; (4) rotated states $|\psi_{4,4'}\rangle = \cos(\pi/8|0\rangle \pm \sin(\pi/8|1\rangle$; (5) phase-modulated states $|\psi_{5,5'}\rangle = \cos(\pi/8)|0\rangle\pm e^{i\pi/3}\sin(\pi/8)|1\rangle$; and (6) asymmetric states $|\psi_6\rangle \propto |0\rangle + 0.5|1\rangle$ and $|\psi_{6'}\rangle\propto |1\rangle - 0.5|0\rangle$.  
		(b) Trace distance $D(t)$ exhibits non-monotonic decay with state-dependent revivals, signaling information backflow for the six orthogonal state pairs with $T=0.2/\omega_0$ and $\omega_c=\lambda=0.1\omega_0$.  
	}\label{fig6}
\end{figure}

    In this appendix, we compute the non-Markovianity of the studied system using the HEOM method.
    To characterize non-Markovian dynamics, we employ the Breuer-Laine-Piilo (BLP)  measure, which quantifies memory effects through information backflow from the environment to the system. The non-Markovianity \(\mathcal{N}\) is defined as the maximal time-integrated increase in trace distance between two quantum states~\cite{PhysRevLett.103.210401}:
    \begin{equation}\label{nonm}
	    \mathcal{N} = \max_{\rho_{1,2}(0)} \int_{\sigma(t)>0} \sigma(t)\, dt, \quad \sigma(t) = \frac{d}{dt} D(t),
    \end{equation}
    where \(D(t) = \frac{1}{2} \text{Tr}|\rho_1(t) - \rho_2(t)|\) denotes the trace distance. A positive \(\sigma(t)\) reflects an increase in state distinguishability, serving as a direct indicator of non-Markovian information backflow.

    To compute \(\mathcal{N}\) accurately, we employ the HEOM method, which captures system–environment correlations beyond perturbative and Markovian approximations. A Drude–Lorentz spectral density is used, with the coupling strength and cutoff frequency determining the degree of non-Markovianity~\cite{PhysRevLett.116.020503}.
    As shown in Eq.~(\ref{nonm}), the evaluation of \(\mathcal{N}\) depends on choosing an optimal pair of initial states \(\rho_1(0)\) and \(\rho_2(0)\). For a two-level system, orthogonal states maximize the initial trace distance and hence \(\mathcal{N}\). However, this simplification may not hold for more complex quantum systems. We systematically examine six representative pairs, as shown in Fig.~\ref{fig6}(a), and identify \(|+\rangle, |-\rangle \) as yielding the largest \(\mathcal{N}\).  This finding is consistent with prior research~\cite{PhysRevA.98.062106}, which employed numerical simulations to explore all possible states, and is further supported by the corresponding trace distance dynamics shown in Fig.~\ref{fig6}(b).  Notably, this particular pair exhibits the most significant and sustained revivals—a hallmark of strong non-Markovian backflow.

	\section{Two-qubit thermometry in a common bath}  \label{appd}
	\begin{figure}[th]
		\centering\includegraphics[width=8.3cm]{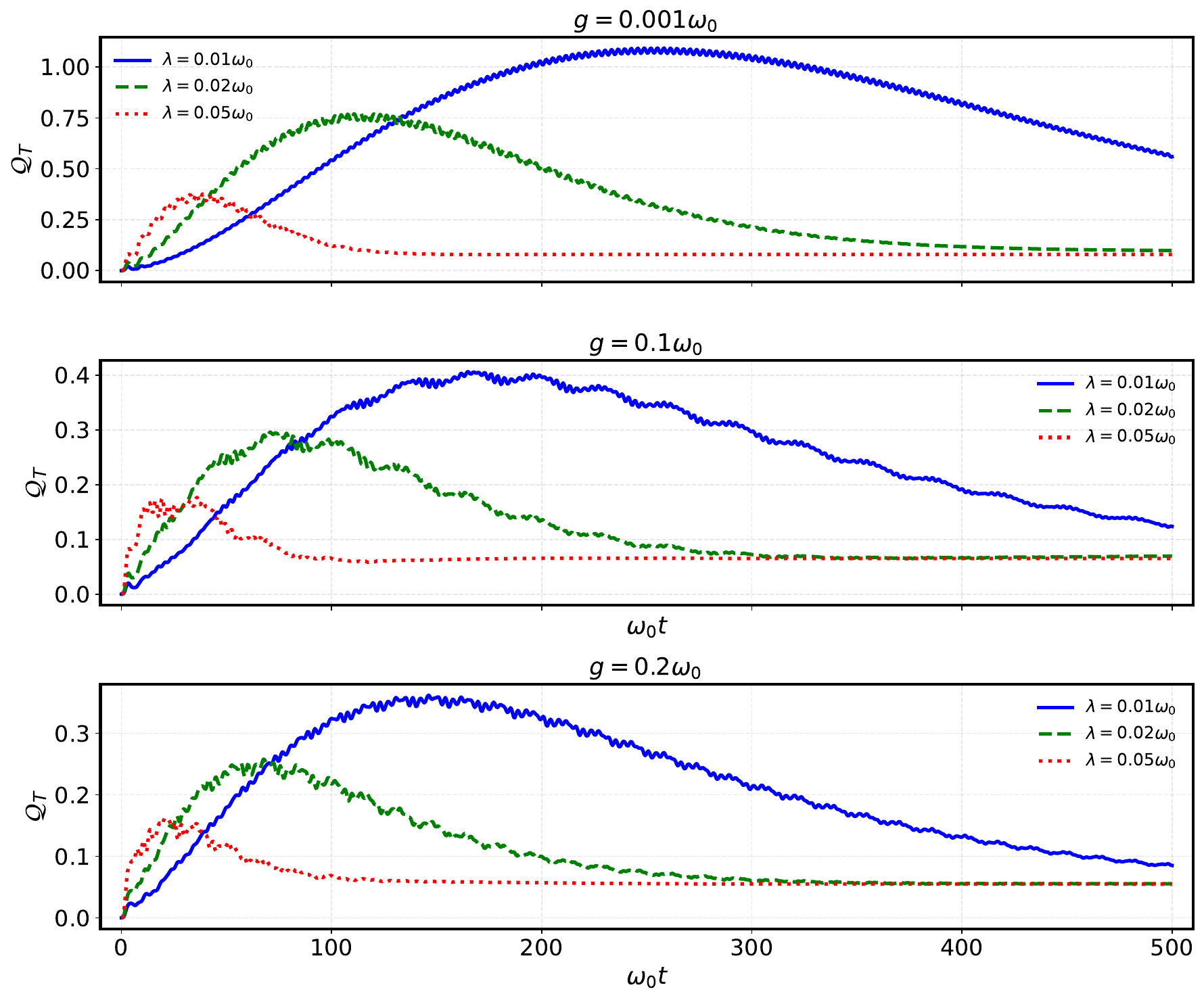}
		\caption{
			QSNR for two qubits coupled to a common bath with varying coupling strengths $g$, where other parameters are fixed at $T=0.2/\omega_0$ and $\omega_c=0.1\omega_0$.
		}\label{fig7}
	\end{figure}
The hierarchical equations of motion (HEOM) framework employed in our study is inherently suitable for addressing multi-qubit quantum thermometry problems. To demonstrate this versatility, we investigate a representative scenario involving two qubits coupled to a common bosonic thermal bath. Our goal is to assess whether the key observations made in the single-qubit case—particularly the sensitivity enhancement in the weak-coupling regime—remain valid in this extended setting.

The two-qubit system is governed by a Hamiltonian of the same general form as Eq.~(\ref{ham0}) in the main text, but with distinct definitions for the system Hamiltonian $\hat{H}_{S}$ and the coupling operator $\hat{S}$. Specifically, the system Hamiltonian is given by
	\begin{equation}
	\hat{H}_{S} = \frac{\omega_0}{2} \left( \hat{\sigma}_z^{(1)} + \hat{\sigma}_z^{(2)} \right)
	+ \frac{g}{2} \left( \hat{\sigma}_x^{(1)}\hat{\sigma}_x^{(2)} + \hat{\sigma}_y^{(1)}\hat{\sigma}_y^{(2)} \right),
\end{equation}
where $g$ denotes the coupling strength between the two qubits.
The system interacts with the environment through the coupling operator
\begin{equation}
	\hat{S} = \hat{\sigma}_x^{(1)} + \hat{\sigma}_x^{(2)}.
\end{equation}

Here, we employ the Drude-Lorentz spectral density and initialize the system in a separable product state:
\begin{equation}
	|\psi_0\rangle = |\psi_1\rangle \otimes |\psi_2\rangle, \quad \text{with} \quad |\psi_{1,2}\rangle = \frac{1}{\sqrt{2}} (|0\rangle + |1\rangle).
\end{equation}
This choice is motivated by prior studies indicating that, for certain thermometric tasks, product states may outperform entangled states due to their superior robustness against decoherence in open quantum systems~\cite{PhysRevResearch.7.023100,PhysRevA.97.012126}.

To quantify temperature sensitivity, we compute the QSNR $\mathcal{Q}_T$ using the spectral decomposition of the density matrix $\rho = \sum_i p_i |\psi_i\rangle\langle\psi_i|$ and its temperature derivative $\partial_T \rho$. The QSNR for a general mixed state is given by
\begin{equation}
	\mathcal{Q}_T = T^2 \sum_{i,j} \frac{2 \left| \langle \psi_i | \partial_T \rho | \psi_j \rangle \right|^2}{p_i + p_j}, \quad \text{for } p_i + p_j \neq 0,
\end{equation}
where $p_i$ and $|\psi_i\rangle$ denote the eigenvalues and eigenvectors of $\rho$, respectively.

The numerical results in Fig.~\ref{fig7} confirm that weak system–environment coupling enhances the dynamical behavior of the QSNR, consistent with the single-qubit case. In the model we consider, a two-qubit thermometer does not necessarily outperform its single-qubit counterpart. In particular, the QSNR remains consistently higher in the weak-coupling regime throughout the evolution. Furthermore, we find that relatively small qubit–qubit interaction strengths (e.g., $g = 0.001\omega_0$) are more favorable for enhancing the QSNR than stronger interactions (e.g., $g = 0.2\omega_0$) within the common-bath model considered here.

These findings suggest that weak qubit–qubit coupling may, in fact, improve thermometric precision in the common-bath scenario considered here. They also highlight the flexibility and scalability of the HEOM method in capturing rich non-Markovian dynamics in multipartite open quantum systems, thereby broadening its applicability in quantum metrology beyond the single-qubit regime.

	\acknowledgments
	Q.S.T. acknowledges support from the National
	Natural Science Foundation of China (NSFC) (12275077).
	X. Xiao is supported by the National Natural Science Foundation of China under Grant Nos. 12265004 and Jiangxi Provincial Natural Science Foundation under Grant No. 20242BAB26010.
	W.W. acknowledges the supports from the National Natural Science Foundation of China (Grant Nos.
	12375015 and 12247101).
	
	\bibliography{reference}
	
\end{document}